\documentstyle[12pt,aasms4]{article}

\input epsf.sty
\input{psfig.tex}

\newcommand{\etal}{et~al.\ }
\newcommand{\eg}{e.g.\ }
\newcommand{\ie}{i.e.\ }
\newcommand{\Msun}{M$_{\odot}$}

\newcommand{\kms}{km~s$^{-1}$}
\newcommand{\CII}{C~{\sc ii}}

\newcommand{\OII}{O~{\sc ii}}

\newcommand{\SiII}{Si~{\sc ii}}
\newcommand{\SiIII}{Si~{\sc iii}}
\newcommand{\SII}{S~{\sc ii}}
\newcommand{\TiII}{Ti~{\sc ii}}

\newcommand{\FeII}{Fe~{\sc ii}}
\newcommand{\FeIII}{Fe~{\sc iii}}

\newcommand{\Nifs}{$^{56}$Ni}

\begin{document}

\title{On the presence of Silicon and Carbon in the pre-maximum spectrum of the
Type Ia SN 1990N} 

\author{Paolo A. Mazzali \\
~~ \\ 
Osservatorio Astronomico \\
Via Tiepolo, 11 \\
I-34131 Trieste \\
Italy }

\begin{abstract} 
The spectrum of the normal Type Ia SN 1990N observed very early on (14 days
before $B$ maximum) was analysed by Fisher \etal (1997), who showed that the
large width and the unusual profile of the strong line near 6000\AA\ can be
reproduced if the line is assumed to be due to \CII\ 6578, 6583\AA\ and if
Carbon is located in a high velocity shell. This line is one of the
characterising features of SNe Ia, and is usually thought to be due to \SiII. A
Monte Carlo spectrum synthesis code was used to investigate this suggestion
further. The result is that if a standard explosion model is used the mass
enclosed in the shell at the required high velocity (25,000--35,000 \kms) is
too small to give rise to a strong \CII\ line. At the same time, removing
Silicon has a negative effect on the synthetic spectrum at other wavelengths,
and removing Carbon from the lower velocity regions near the photosphere makes
it difficult to reproduce two weak lines which are naturally explained as \CII,
one of them being the line which Fisher \etal (1997)  suggested is responsible
for the strong 6000\AA\ feature. However, synthetic spectra confirm that
although \SiII\ can reproduce most of the observed 6000\AA\ line, the red wing
of the line extends too far to be compatible with a \SiII\ origin, and that the 
flat bottom of the line is also not easy to reproduce. The best fit is obtained 
for a normal SN~Ia  abundance mix at velocities near the photosphere 
(15,500-19,000 \kms) and an outer Carbon-Silicon shell beyond 20,000 \kms.  
This suggests that mixing is not complete in the outer ejecta of a SN~Ia.  
Observations at even earlier epochs might reveal to what extent a Carbon shell 
is unmixed.   
\end{abstract}


\section{Introduction} 

The distinguishing spectral feature of SNe~Ia is the \SiII\ 6347, 6371\AA\
line. This line is strong in all photospheric-epoch spectra of SNe~Ia,
appearing near 6000\AA\ from before maximum to at least one month after
maximum. Its strength suggested that a large fraction of the material ejected
in a SN~Ia explosion is in the form of Si and other so-called 'Intermediate
Mass Elements' (\ie between CO and the Fe-group). These elements (mostly Si, S,
and Ca) are synthesised in regions where thermonuclear burning of the
progenitor Carbon-Oxygen white dwarf (WD) is incomplete, not leading to the
production of \Nifs\ (Nomoto \etal 1984). Since this is only possible if the
burning wave that disrupts the WD is subsonic (aka a deflagration wave), the
identification of this line was an important step in understanding that burning
must must proceed subsonically in a significant fraction of the WD. Explosion
models based on this assumption could successfully reproduce the light curve
and spectra of SNe~Ia starting from a Chandrasekhar mass CO WD (\eg the
deflagration model W7, Nomoto \etal 1984, Branch \etal 1985; or various delayed
detonation and pulsational deflagration models presented by Woosley \& Weaver
(1994) and Woosley (1997).

Fisher \etal (1997) (hereafter FBNB) re-analysed the earliest spectrum ever
recorded of a SN~Ia, that of SN~1990N on 26 June 1990, which is as early as 14
days before maximum $B$ light (Leibundgut \etal 1991). Using a simple but
powerful parametrised LTE code, FBNB reproduced the UV-optical spectrum
satisfactorily, but noticed that the broad and flat-bottomed profile of the
'\SiII' line, observed near 6040\AA, could not be reproduced using Si and an
exponential radial dependence of the line optical depth. Other authors (Jeffery
\etal 1992, Mazzali \etal 1993) encountered essentially the same problem, even
though they used more sophisticated codes. As an alternative solution, FBNB
showed that the line profile could be successfully modelled by a high velocity
Carbon shell ($26,000 \leq v \leq 35,000$ \kms). This proposed alternative
identification is not in contradiction with the basic properties of successful
explosion models, which usually have a layer of unburned progenitor material
(CO in this case) at the top of the ejecta and hence expanding at the highest
velocities. Such a layer would be most easily detectable for its effect on the
spectrum when it is not yet very far removed from the momentary and receding
photosphere, and therefore it is natural to expect that the very early spectrum
of the normal SN~Ia 1990N would be one of the best places to find evidence for
it.

This suggestion has very interesting consequences, both because it sets some
limit to the mixing taking place in the highest velocity part of the ejecta and
because it points out a possible source of error in measures of very early
spectra of SNe~Ia, especially if the spectra are of poorer quality than that of
SN~1990N, a typical situation for SNe at high redshift. 

Independently of FBNB, we have been modelling the same UV-optical spectrum of
SN~1990N in an effort to determine its epoch as accurately as possible so as to
set some limits on the rise time to maximum, in view of the current debate
about this parameter (Mazzali \& Schmidt 2000, in prep.). We have used a Monte
Carlo (MC) code based on that described in Mazzali \& Lucy (1993), but modified
to include an extended line list and photon branching (Lucy 1999, Mazzali
2000). Although we were trying to obtain a good overall fit of the spectrum,
and did not pay much attention to small features, we noticed that in just about
every synthetic spectrum we computed two rather weak features were always well
reproduced as \CII\ lines: a small absorption near 6350\AA, which is attributed
to \CII\ 6578, 6583\AA, and the absorption near 6900\AA, attributed to \CII\
7231, 7236\AA\ with a contribution from \OII\ 7321\AA\ to make the feature
relatively broad. The strong absorption line near 6040\AA\ was matched by the
\SiII\ line, but the line profile was not reproduced correctly, as in all
previous work. In all models the abundances were homogeneous above the
photosphere. The model we finally selected had an epoch $t = 5$ days (implying
$t(Max) = 19$ days), distance modulus $\mu = 32.00$ mag, luminosity $L = 2.30
\cdot 10^{42}$ erg and photospheric velocity $v_{ph} = 15750$ \kms. The
synthetic spectrum is shown as the thin continuous line in Fig. 1. The 
velocity indicated by the two weak features in the red, if they are interpreted
as \CII\ lines, is about 12000 \kms, which is somewhat smaller than $v_{ph}$. 
In fact the synthetic \CII\ lines, especially that at 6350\AA, fall at a
somewhat shorter wavelength than the observed ones. Nevertheless, the
coincidence was striking.

Thus we turned our attention to the FBNB paper. FBNB show fits to the entire
spectrum using different compositions in their Fig.1, which was unfortunately
printed too small to verify whether the two small features discussed above are
reproduced as \CII. Nevertheless, in their Fig.2, where they show two different
ways to reproduce the 6040\AA\ feature - using Si and C, respectively, it is
clear that neither the 6350\AA\ feature nor the one at 6900\AA\ are reproduced 
with the `high velocity C shell' model. On the contrary, while in that model
the \CII\ 6578, 6583\AA\ doublet gives rise to the strong and flat-bottomed
6040\AA\ absorption, a weaker, broad and also flat-bottomed absorption is also
visible near 6500\AA, which is not present in the observed spectrum. This is
exactly where the \CII\ 7321, 7236\AA\ doublet would fall if Carbon were
located in a shell between 26,000 and 35,000 \kms. The ratio of the equivalent
widths of the two features is large, with a value of at least 10, which is
larger than the observed ratio but may result from the assumption of LTE since
the redder doublet comes from a more highly excited level (16.33 v. 14.45 eV).
We therefore considered it worthwhile to tackle once more the problem of what
might give rise to the observed 6040\AA\ line using synthetic spectra. In the
next sections we discuss various alternatives.

\section{Can Carbon replace Silicon?} 

We have shown that a fully mixed model reproduces two weak \CII\ features, and
gives a less-than-satisfactory fit of the 6040\AA\ feature as \SiII. If we are
to explain this feature as due to \CII\ instead, the first necessary step is to
remove Silicon from the mixture (as FBNB did in the central panel of their
Fig.1) and ascertain what influence this has on the synthetic spectrum. In
Fig.1 we show two models obtained for a homogeneous composition: the dashed
line is for a full W7 composition mix, including both Si and C, as discussed in
the previous section.  The peak at 3500\AA, which is not reproduced by our
synthetic spectrum, is in the MMT data. Since the calibration at the edge of
the spectrum is tricky, the poor fit at this wavelength does not necessarily
mean that the model is wrong.

The dotted line in Fig.1 is a model where Silicon has been removed and its
abundance (0.2 by mass) is assigned to Carbon instead. In this model both \CII\
lines, but especially the 6580\AA\ one, are too strong, and are not compatible
with the observations. Removing Si, and not placing C in a high velocity shell,
clearly destroys the fit of the 6040\AA\ feature. Furthermore, removing Si has
negative consequences elsewhere in the spectrum, which cannot be remedied by
introducing C at high velocities: 1) the weak \SiIII\ 4553, 4568\AA\ line,
observed near 4400\AA, a typical feature of all but the coolest SNe~Ia before
and near maximum, is now lost; 2) the strong absorption near 4800\AA\ has about
equal contributions from \SiII\ 5041, 5056\AA\ (falling near 4750\AA) and
\FeIII\ 5156 + \FeII\ 5169 \AA\ (falling near 4850\AA) in the spectrum obtained
for a W7 mix: when Si is removed, only the red part of the absorption is left,
and this absorption is too weak to cause a strong re-emission peak, which is
observed near 5100\AA; 3) the shape of the spectrum near 5500\AA\ is influenced
mostly by \SiIII\ 5740\AA\ (near 5425\AA) and \SiII\ 5979\AA\ (near 5600\AA),
although the broader absorption  extending to 5200\AA\ is due to several lines
of \SII. When Si is removed from the mixture, only the \SII\ lines are left and
the absorption in the model is too far to the blue. These three points show
clearly that Silicon cannot be completely removed from the relatively high
velocity regions near the photosphere. But if Si is allowed in the mix, then
the doublet 6347, 6371\AA\ is the strongest Si line. Therefore, it appears
unavoidable to conclude that \SiII\ contributes significantly to the 6040\AA\
feature even at this early epoch.

\section{A Carbon-shell model?} 

FBNB could fit the 6040\AA\ feature placing Carbon at high velocity, but since
they used a parametrised model they did not quantify the mass of C required to
obtain such a strong synthetic line. Since the \CII\ lines are weak even in a
fully mixed model, this is an interesting numerical experiment. The model shown
as a dotted line in Fig.1, where Si had been replaced by C, had a Carbon
abundance of 0.25 by mass. Since the total mass above the photosphere in this
model is 0.072\Msun, the C mass was 0.018\Msun, and the strength of the \CII\
6578, 6583\AA\ line was comparable to that of the observed 6040\AA\ feature,
but the velocity was wrong. The velocity can be reconciled if C is enhanced
only for velocities larger than  $v_{sh} = 26000$\kms, as FBNB suggested. In
our next synthetic spectrum we replaced Si with C only above this velocity,
while between $v_{ph}$ and $v_{sh}$ we just redistributed the Si abundance
among all elements, proportionally to their respective abundances (the
composition is dominated by Oxygen). Because of the steeply falling density
($\rho \propto r^{-7}$), when C is confined to very high velocities its mass is
reduced. This reduces the C mass to only $10^{-3}$\Msun. The result is shown in
Fig.2 (thin continuous line): both \CII\ lines are so weak that they do not
produce a noticeable feature in the spectrum. This calculation reproduces all
essential features of the FBNB 'high velocity C shell' model, and at the same
time is consistent with the hydrodynamical model of a SN~Ia: the result is that
not only the synthetic spectrum does not reproduce the strong 6040\AA\ feature,
but the absence of Si creates the three problems listed in Sect.2, and the
displacement of C in velocity space means that the two weak features which are
attributed to \CII\ are not reproduced either.

Obviously, it must be possible to fit the 6040\AA\ feature as \CII\ by
increasing the \CII\ line opacity in the high velocity shells. We show this as
the dashed line in Fig.2. The values by which we had to multiply the Sobolev
optical depth to obtain this spectrum are an increasing function of velocity,
ranging from 10 at 22500\kms\ to $10^6$ at 30000\kms, the outer limit of the
assumed ejecta distribution. Although these values do give a very good fit to
the observed feature, they really have no physical basis. If the increased
optical depth is attributed to the number of absorbing \CII\ ions, this would
lead to a completely unrealistic C mass, more than 1\Msun, at these large
velocities. Furthermore, all the problems raised by the absence of Si remain,
and the two small features are not fitted at all.

It is interesting to note that although we increased the optical depth for all
\CII\ lines, we obtained a strong 6578, 6583\AA\ line, but not a strong 7231,
7236\AA\ line. This is because the latter doublet comes from a more highly
excited level, whose population falls more steeply with radius, and so even
larger factors would be required to enhance the line at very high velocities.

This test confirms that the 6040\AA\ feature must be at least predominantly due
to \SiII, and that placing Carbon in a high velocity shell does not give
sufficient opacity, even if C dominates the composition.

\section{A Silicon model?} 

We believe we have provided ample evidence that the 6040\AA\ feature cannot be
entirely - or even mostly - due to \CII\ lines. As an alternative solution, can
the distribution of Silicon be modified so that a better fit to the line can be
obtained? If we look back at Fig.1 we notice that the profile of the \SiII\
6347, 6371\AA\ line is too sharp, and that absorption is missing both in the
blue (between 5800 and 5900\AA) and in the red (between 6100 and 6200\AA).
These regions correspond to Si velocities of about 23000 and 10000km/s,
respectively. Therefore, we can expect that we can produce more blue absorption
by increasing Si at high velocity, but the velocity required to fit the red
part of the absorption is significantly smaller than that of the photosphere.

In Fig. 3 we show a model where the Si abundance has been increased by a factor
of 3 above 23000\kms, at the expense of all other elements. This gives a
reasonably good fit to the blue side of the line. Other \SiII\ lines are
weaker, so they are not much affected by this change, and the rest of the
spectrum is not very different from that of Fig.1. As regards the red part of
the line though, setting the photospheric velocity at 15750 \kms, imposes the 
constraint that the reddest wavelength of \SiII\ is about 6050\AA. The velocity
of the photosphere cannot be reduced to about 10000 \kms\ without changing the
aspect of the synthetic spectrum completely (see Mazzali \& Schmidt 2000, in
prep.). Therefore we agree with FBNB that Si alone is unable to explain the
observed width and profile of the 6040\AA\ feature.

\section{A possible solution: Silicon {\em and} Carbon}

On the basis of the models presented above we suggest that Si is responsible
for most of the 6040\AA\ feature in SN~1990N, and in particular for its
extended blue side, but an alternative origin must be found for the red part of
the line. At the same time, we have shown that the small feature near 6300\AA\
is most likely due to \CII\ at a near-photospheric velocity. Our line list,
which was derived from that of Kurucz \& Bell (1995) does not seem to offer any
reasonably strong line with a rest wavelength of about 6450\AA, which would
naturally explain the red extension of the 6040\AA\ feature by redshifting at
the velocity of the photosphere. The nearest strong lines to the red of the
\SiII\ doublet are indeed the \CII\ ones. A Carbon shell centred at about
20000\kms\ might explain the observations. The questions are can C be
responsible for both the small absorption at 6300\AA\ (via its
near-photospheric distribution) and the red wing at $\sim 6100$\AA\ (via an
enhanced C abundance at $v \sim 20000$\kms), and if so is the required C
enhancement compatible with the hydrodinamical structure of a SN~Ia?

The answer is shown in Fig.4. This is a model computed with the same parameters
as the previous ones and the following distributions: Carbon: 0.05 by mass up
to $v = 18500$\kms, then increasing smoothly and reaching a peak value 0.50
above $v = 21500$\kms; Silicon: 0.20 by mass up to $v = 21500$\kms, and then
increasing to 0.5 above $v = 22500$\kms. The integration is limited to an outer
velocity of 30000\kms. 

The model reproduces the observed profile reasonably well. The observed flat
bottom of the line may require more ad hoc adjustments of the distribution of
the elements. The `average' positions of the near-photospheric and of the high
velocity components of the synthetic \CII\ 6578, 6583\AA\ absorption are 
marked in Fig.4. The near-photospheric component falls at a longer wavelength
than the the observed 6300\AA\ feature. The difference in velocity is about
3600\kms. If the adopted photospheric velocity is correct, this offest is a
puzzle. We have no alternative identification other than \CII\ to offer for
that feature, since our line list offers only very weak \TiII\ lines with rest
wavelength of about 6700\AA. But even if the 6300\AA\ feature is not \CII, our
conclusion regarding the origin of the 6040\AA\ feature does not change.  

The scenario we are faced with is then the following: a mixed W7 composition 
(C=0.05 and Si=0.20 by mass, respectively) holds out to $v \sim 19000$ kms, 
but outside that velocity a C-Si shell develops. The near-photospheric C
abundance may be lower if the line at 6300\AA\ is not \CII.  With the exception
of the Si shell, this is consistent with incomplete mixing in the SN ejecta.
Maybe \SII\ lines are responsible for the extra absorption between 5800 and
5900\AA, but our line data do not support that.

\section{Conclusions} 

We have shown that the strong and broad absorption near 6040\AA\ in the d -14
spectrum of SN~1990N must be predominantly due to \SiII\ 6347, 6371\AA. If we
reproduce the line as \CII, eliminating Si from the mixture as suggested by
FBNB, other regions of the spectrum where Si lines are strong are not well
reproduced. Also, the mass enclosed in the high velocity shell which could give
rise to a blueshifted \CII\ line is very small, so that the synthetic line is
much too weak, and the observed profile can only be reproduced if the optical
depth of the \CII\ lines is increased by unrealistic factors (up to $10^6$). 
On the other hand, we have confirmed the result of FBNB that \SiII\ alone
cannot reproduce the entire feature because the red edge of the line has a
blueshift much smaller than that of the \SiII\ line at the velocity of the
momentary photosphere. 

As a possible contribution for that part of the feature we also suggest \CII\
in a high velocity shell, but unlike FBNB we enhance C in a shell between
19000 and 30000\kms, which contains a small (0.04\Msun), but nevertheless large
enough mass to give rise to a rather strong line without resorting to
artificially large departure coefficients to increase the line optical depth.
The increased C abundance in this outer shell does not significantly affect
the total mass of C in the ejecta. At the same time, mixed C at
near-photospheric velocities gives rise to two weak synthetic absorptions at
6250 and 6900\AA, which are a reasonable match for two observed features at
6300 and 6900\AA. The blending of the \SiII\ and \CII\ lines may give rise to
the flat bottom of the observed 6040\AA\ feature, although our model does not
reproduce that.

The implications of our result are that an outer zone where Carbon is not fully
mixed does indeed appear to exist. This zone is narrow, and contains only a
small mass, so the photosphere passes through it rapidly. Our models suggest
that Si is also present there. Therefore, observing a SN~Ia very early on, when
the photosphere is still at $v \sim 20000$\kms, would be very interesting. If
both Si and C are present at high velocities, we do not expect that the feature
should appear very different from what is observed in the spectrum of SN~1990N
we have analysed in this paper. Alternatively, if only C were present, the
`\SiII' absorption would be much redder ($\lambda \sim 6150$\AA) and caused
mostly by \CII. This would require observations very soon after the explosion,
roughly 16 days before maximum, which this may not be so very difficult with
SNe at high redshift, whose observed evolution is slowed down by the factor
($1+z$). Such observations would be extremely interesting because they would
probe the outermost part of the ejecta and could thus clarify by direct
observation to what extent mixing actually takes place. Finally, our findings
must serve as a caveat against deriving SN~Ia properties from the `\SiII' line
if the SN is observed very early on.

\acknowledgements{ It is a pleasure to thank Brian Schmidt for providing
the observed spectrum in a form a modeller can easily work with. }

\bibliographystyle{mn}


\begin{figure}[t]
\epsfxsize=13.5cm 
\hspace{3.5cm}\epsfbox{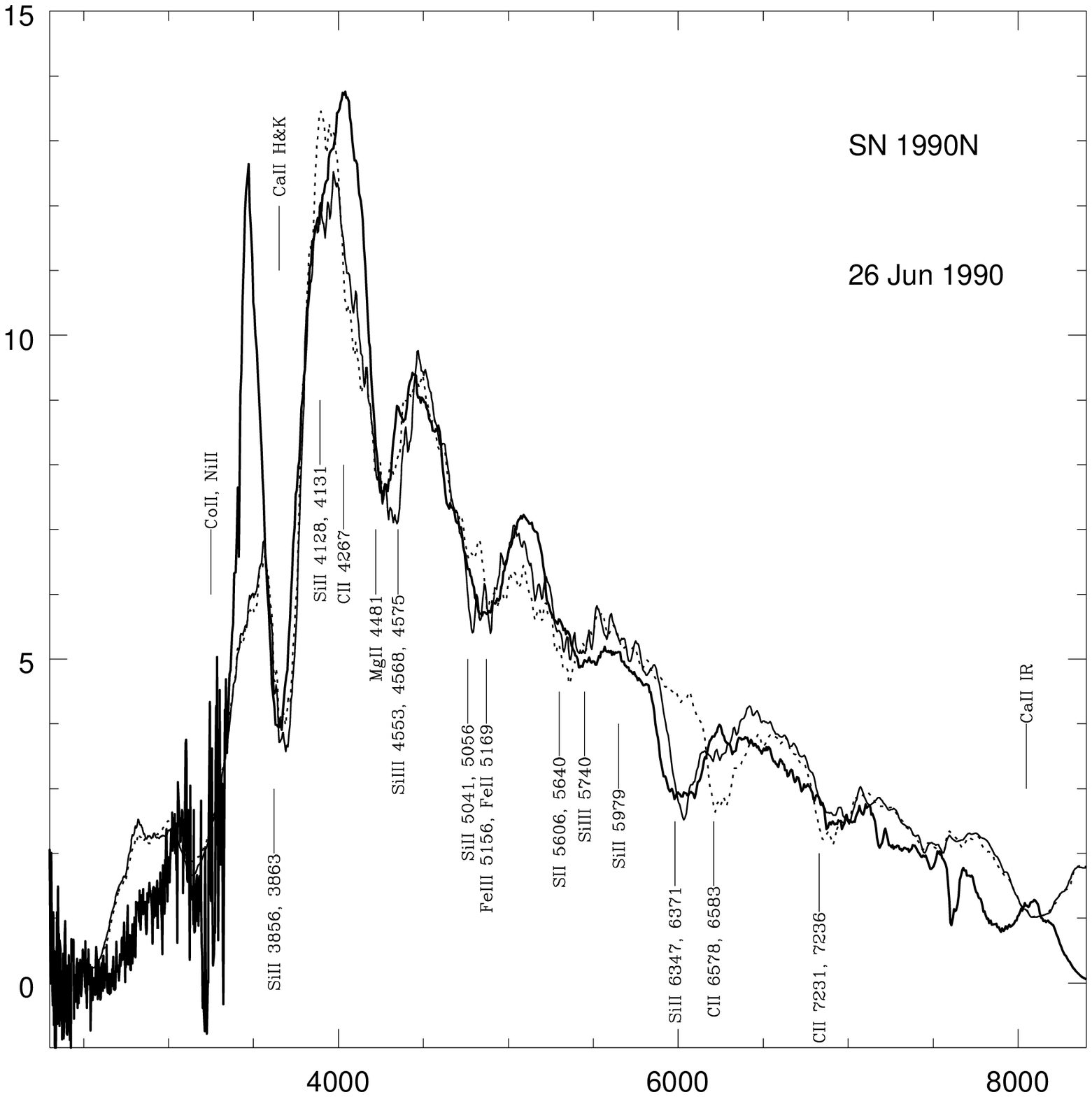}
\caption[h]{The $-14$d UV-optical spectrum of SN~1990N (thick solid line) is
compared to: 
a) a synthetic spectrum computed with a W7-mixed composition (thin solid line), 
which reproduces the overall spectrum, including several \SiII\ and \SiIII\ 
features. The \SiII\ 6347, 6371\AA\ line is too narrow, and two \CII\ lines are 
seen: one, near 6250\AA, may be compatible with the feature at 6300\AA, while 
the other is a good match for the line at 6900\AA; 
b) a synthetic spectrum where Si has been replaced by C throughout the envelope 
(dotted line). The \CII\ lines (especially the one at 6250\AA) are too strong, 
while many Si features are missing. }
\end{figure}

\begin{figure}[t]
\epsfxsize=13.5cm 
\hspace{3.5cm}\epsfbox{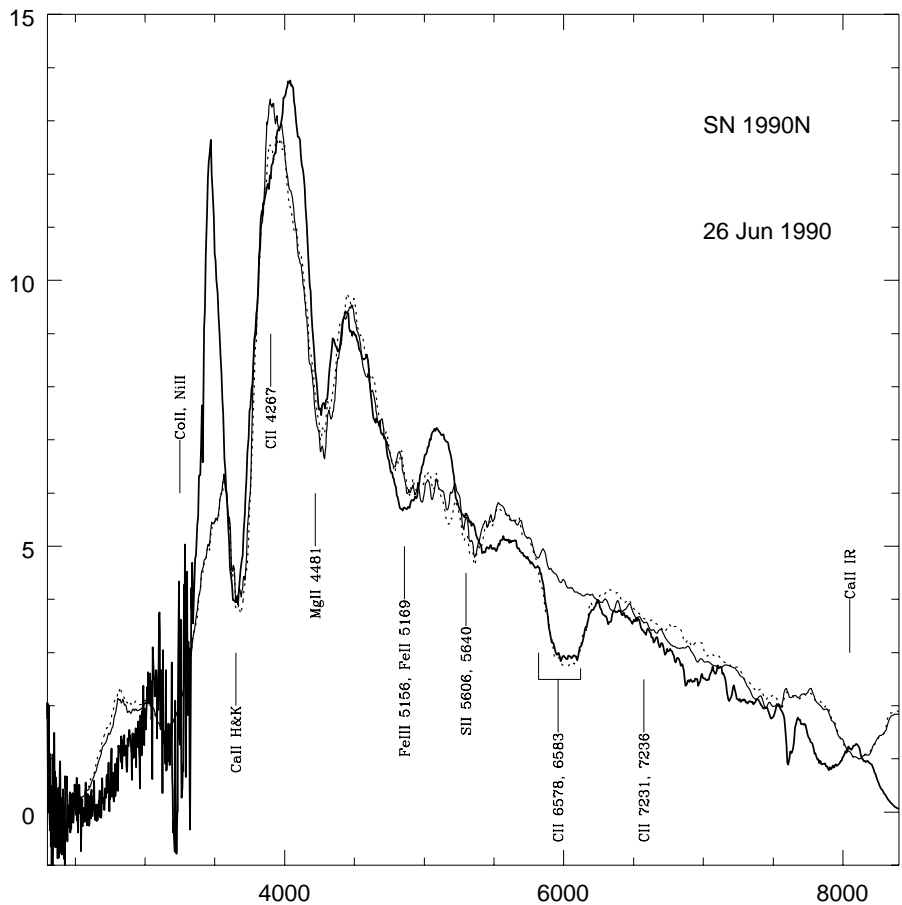}
\caption[h]{The $-14$d UV-optical spectrum of SN~1990N (thick solid line) is
compared to: 
a) a synthetic spectrum where Si is eliminated, but it is replaced by C only at 
$v \geq 26000$\kms, in analogy with FBNB (thin solid line). The \CII\ lines are 
too weak to leave a signature in the synthetic spectrum; 
b) a synthetic spectrum similar to the one above, but with artificially
increased \CII\ line opacities. This fits the 6040\AA\ line, but fails to
reproduce other \SiII\ and \SiIII\ lines and the two weak lines at 6300 and
6900\AA. }
\end{figure}

\begin{figure}[t]
\epsfxsize=13.5cm 
\hspace{3.5cm}\epsfbox{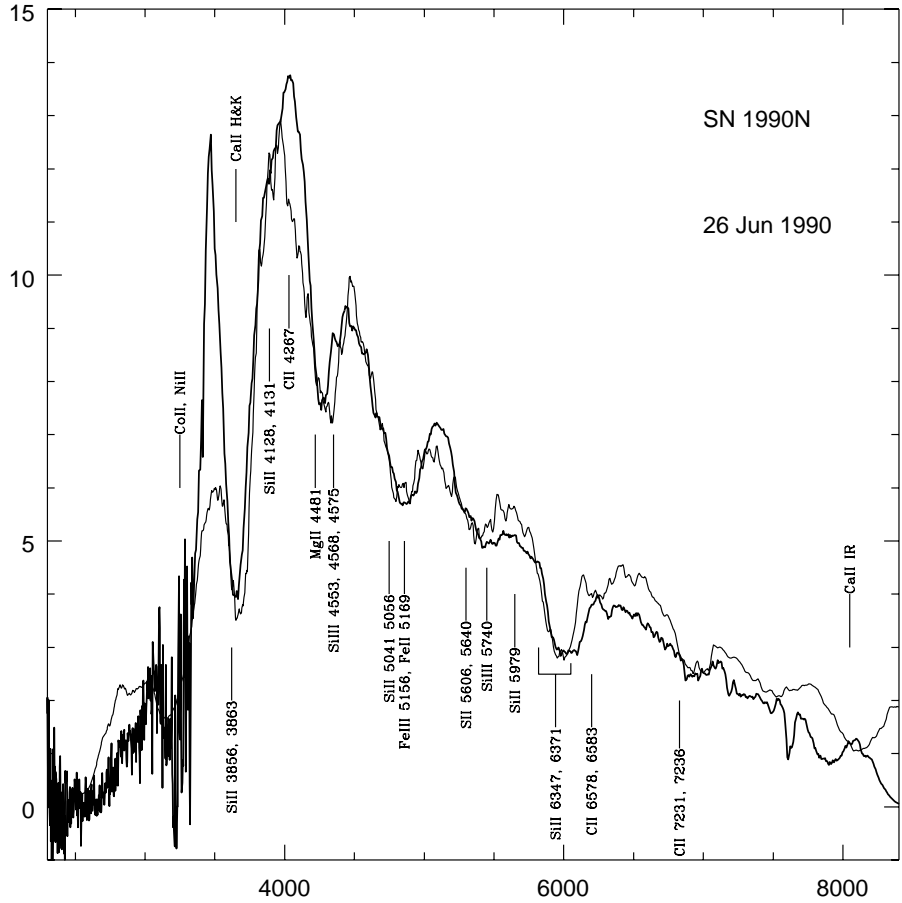}
\caption[h]{The $-14$d UV-optical spectrum of SN~1990N (thick solid line) and a
synthetic spectrum computed for an increased Si abundance above
23000\kms (thin solid line). This reproduces the blue side of the 6040\AA\ 
line, but not the red side. }
\end{figure}

\begin{figure}[t]
\epsfxsize=13.5cm 
\hspace{3.5cm}\epsfbox{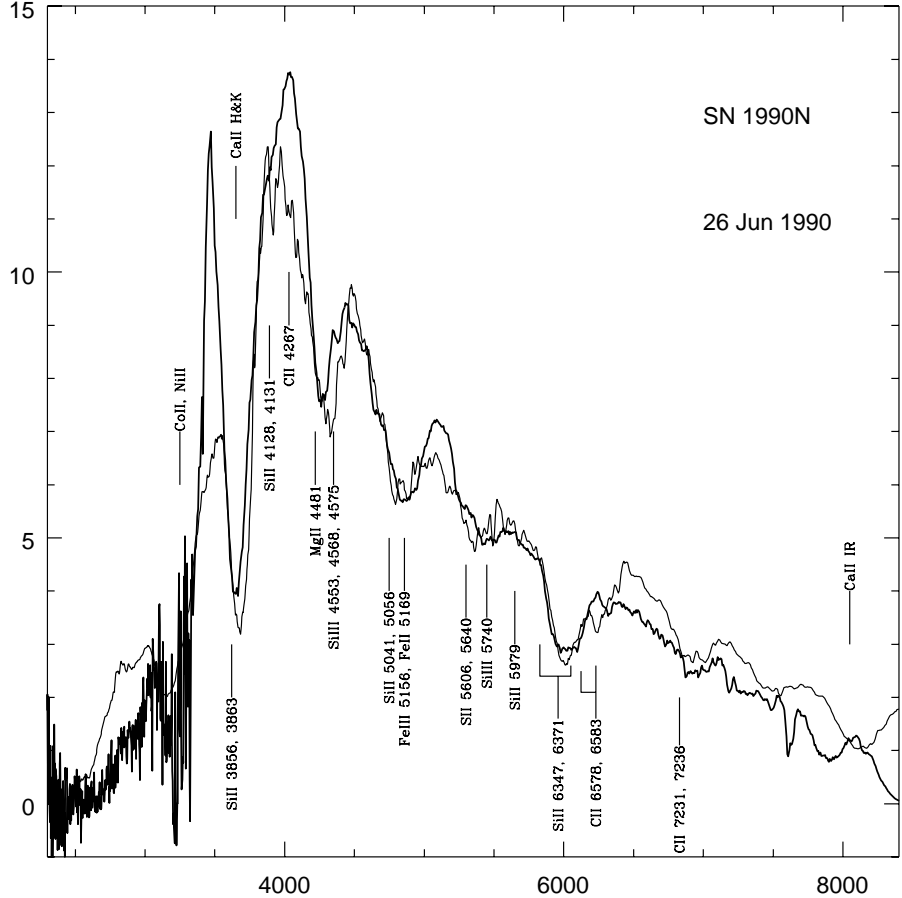}
\caption[h]{The $-14$d UV-optical spectrum of SN~1990N (thick solid line) and 
a synthetic spectrum computed with increased Si and C abundances above about
20000\kms\ (see text for details). This reproduces most observed features,
including the broad 6040\AA\ line and the two weak \CII\ lines, although the
near-photospheric component of the 6578, 6583\AA\ line does not match the
wavelength of the observed 6300\AA\ absorption. }
\end{figure}

\end{document}